\documentclass[journal=langd5,manuscript=article,layout=twocolumn]{achemso}

\usepackage{textcomp}  % used in |textcelsius{}

\author{Julyan H. E. Cartwright}
\affiliation[Instituto Andaluz de Ciencias de la Tierra (IACT), CSIC--Universidad de Granada]
{Instituto Andaluz de Ciencias de la Tierra (IACT), CSIC--Universidad de Granada, Armilla, Spain}

\author{Bruno Escribano}
\email{bescribano@bcamath.org}
\affiliation[Basque Center for Applied Mathematics (BCAM)]
{Basque Center for Applied Mathematics (BCAM), Bilbao, Spain}

\author{Diego L. Gonz\'alez}
\affiliation[Istituto per la Microelettronica e i Microsistemi (IMM), CNR]
{Istituto per la Microelettronica e i Microsistemi (IMM), CNR, Bologna, Italy}

\author{C. Ignacio Sainz-D\'{\i}az}
\affiliation[Instituto Andaluz de Ciencias de la Tierra (IACT), CSIC--Universidad de Granada]
{Instituto Andaluz de Ciencias de la Tierra (IACT), CSIC--Universidad de Granada, Armilla, Spain}

\author{Idan Tuval}
\affiliation[Mediterranean Institute for Advanced Studies (IMEDEA), CSIC--Universitat de les Illes Balears]
{Mediterranean Institute for Advanced Studies (IMEDEA), CSIC--Universitat de les Illes Balears, Mallorca, Spain}

\title{Brinicles as a case of inverse chemical gardens}

\begin{document}

\begin{abstract}
Brinicles are hollow tubes of ice from centimetres to metres in length that form under floating sea ice in the polar oceans when dense, cold brine drains downwards from sea ice into sea water close to its freezing point. When this extremely cold brine leaves the ice it freezes the water it comes into contact with; a hollow tube of ice --- a brinicle --- growing downwards around the plume of descending brine. We show that brinicles can be understood as a form of the self-assembled tubular precipitation structures termed chemical gardens, plant-like structures formed on placing together a soluble metal salt, often in the form of a seed crystal, and an aqueous solution of one of many anions, often silicate. On one hand, in the case of classical chemical gardens, an osmotic pressure difference across a semipermeable precipitation membrane that filters solutions by rejecting the solute leads to an inflow of water and to its rupture. The internal solution, generally being lighter than the external solution, flows 
up through the break, and as it does so a tube grows upwards by precipitation around the jet of internal solution. 
%Flow continues through this self-organized pump mechanism as long as the osmotic pressure difference across the membrane is maintained by the dissolution of the original metal salt. 
Such chemical-garden tubes can grow to many centimetres in length. In the case of brinicles, on the other hand, in floating sea ice we have porous ice in a mushy layer that filters out water, by freezing it, and allows concentrated brine through. Again there is an osmotic pressure difference leading to a continuing ingress of sea water in a siphon pump mechanism that is sustained as long as the ice continues to freeze. Since the brine that is pumped out is denser than the sea water, and descends rather rises, a brinicle is a downwards growing tube of ice; an inverse chemical garden.
\end{abstract}

%%%%%%%%%%%%%%%%%%%%%%%%%%%%%%%%%%%%%%%%%%%%%%%%%%%%%%%%%%%%%%%%%%%%%
%% Start the main part of the manuscript here.
%%%%%%%%%%%%%%%%%%%%%%%%%%%%%%%%%%%%%%%%%%%%%%%%%%%%%%%%%%%%%%%%%%%%%
\section{Introduction}
\label{}

\begin{figure}[t]
\vspace*{2mm}
\begin{center}
\includegraphics*[width=\columnwidth]{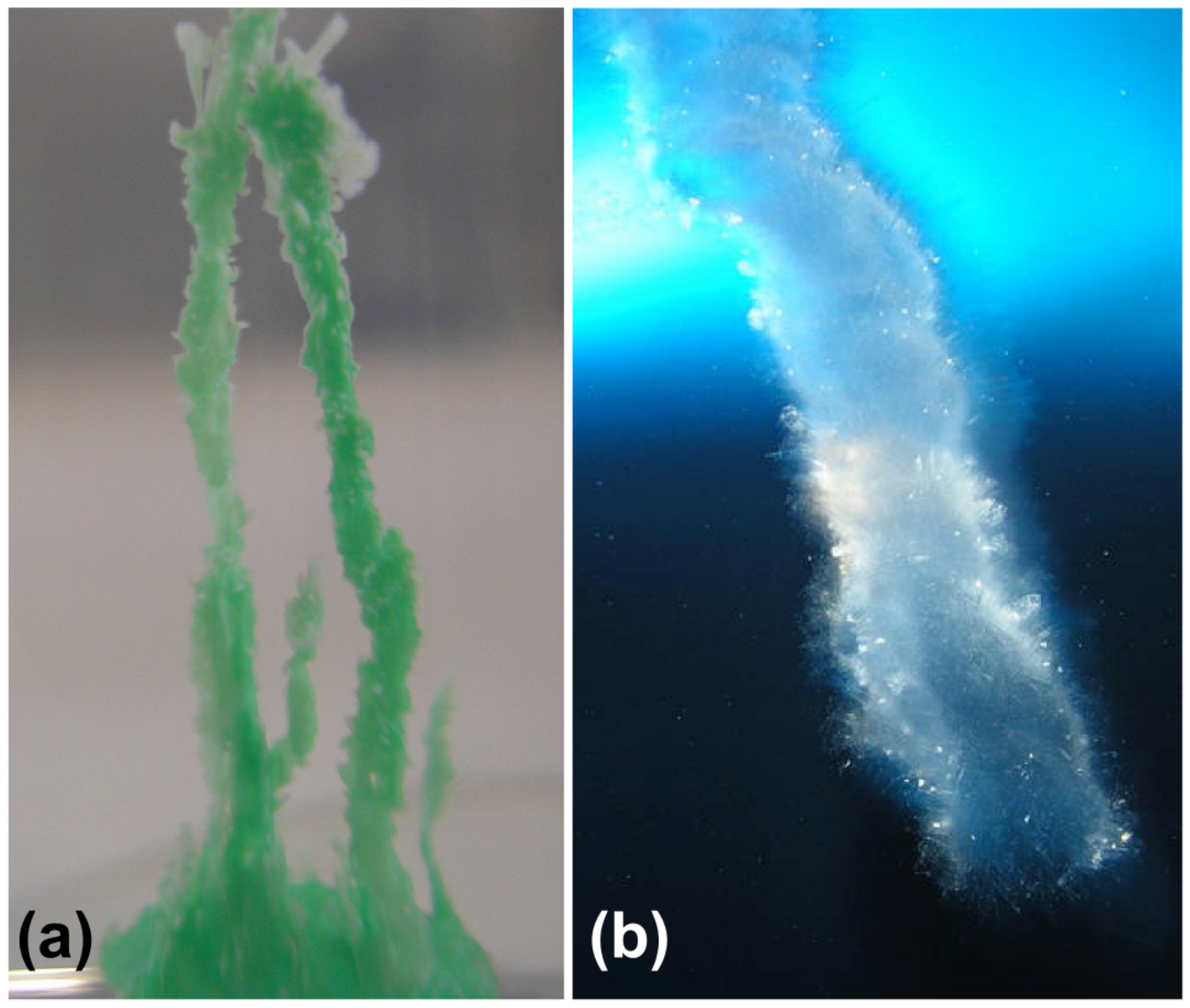}
\end{center}
\caption{
(a) An example of chemical gardens grown from nickel sulphate crystals in a sodium silicate solution. Typically 1--10 cm long.
(b) A brinicle off Ross island, Antarctica. Typically 10--100 cm long. Photography by Rob Robbins on the GOLF 4-3-9 Antarctica Expedition 2010. Image archived by EarthRef.org \cite{earthref}.
}
\label{fig:brinicle}
\end{figure}

Chemical gardens (Fig.~\ref{fig:brinicle}a) are tubular structures that are formed when a metal-salt crystal is immersed in a solution of silicate or other anions \cite{coatman,cartwright,chemgarden1}.
As the metal-ion salt starts to dissolve, it forms a semipermeable membrane about itself. 
The difference of osmotic pressure on the two sides of this semipermeable membrane forces water molecules to pass from the more dilute silicate solution outside to the more concentrated metal-ion solution inside, forming an osmotic pump.
The flow of water molecules inflates the membrane until it ruptures, expelling a jet of metal-ion solution. 
When the metal-ion solution enters in contact with the alkaline silicate solution it precipitates forming a tube around the jet. Thus a chemical garden combines two aspects --- the osmotic pump and tube formation --- which may be found separately in other phenomena, and, more rarely, combined in the same fashion. Examples of such chemical-garden systems outside the laboratory are to be found in cement hydration \cite{double} and in corrosion processes \cite{fontana}.

There are several geological structures that grow in a similar way to chemical gardens. 
Examples include hydrothermal vents \cite{corliss1979}, soda straws \cite{hill1997}, and mud volcanoes \cite{alvisi}.
All these tubular geological formations are a consequence of physical and chemical interactions that combine a membrane or other sort of filtering mechanism and tubular precipitation, solidification, or sedimentation about a fluid flow.

Brinicles (Fig.~\ref{fig:brinicle}b) --- tubes of ice that are found under the pack ice of the Arctic and Antarctic oceans --- are less-well-known examples of tubular patterns in geology.
Brinicles grow around cold streamers of brine beneath sea ice and their size ranges from a few centimetres up to a few metres in length. 
They have also been termed sea-ice stalactites \cite{paige1970, dayton1971,martin1974}, but since an icicle is the icy form of a stalactite, and both stalactites and icicles grow in a different fashion to these structures \cite{neufeld2010,chen2011}, we prefer the term brinicle. 

\begin{figure}[t]
\vspace*{2mm}
\begin{center}
\includegraphics*[width=\columnwidth]{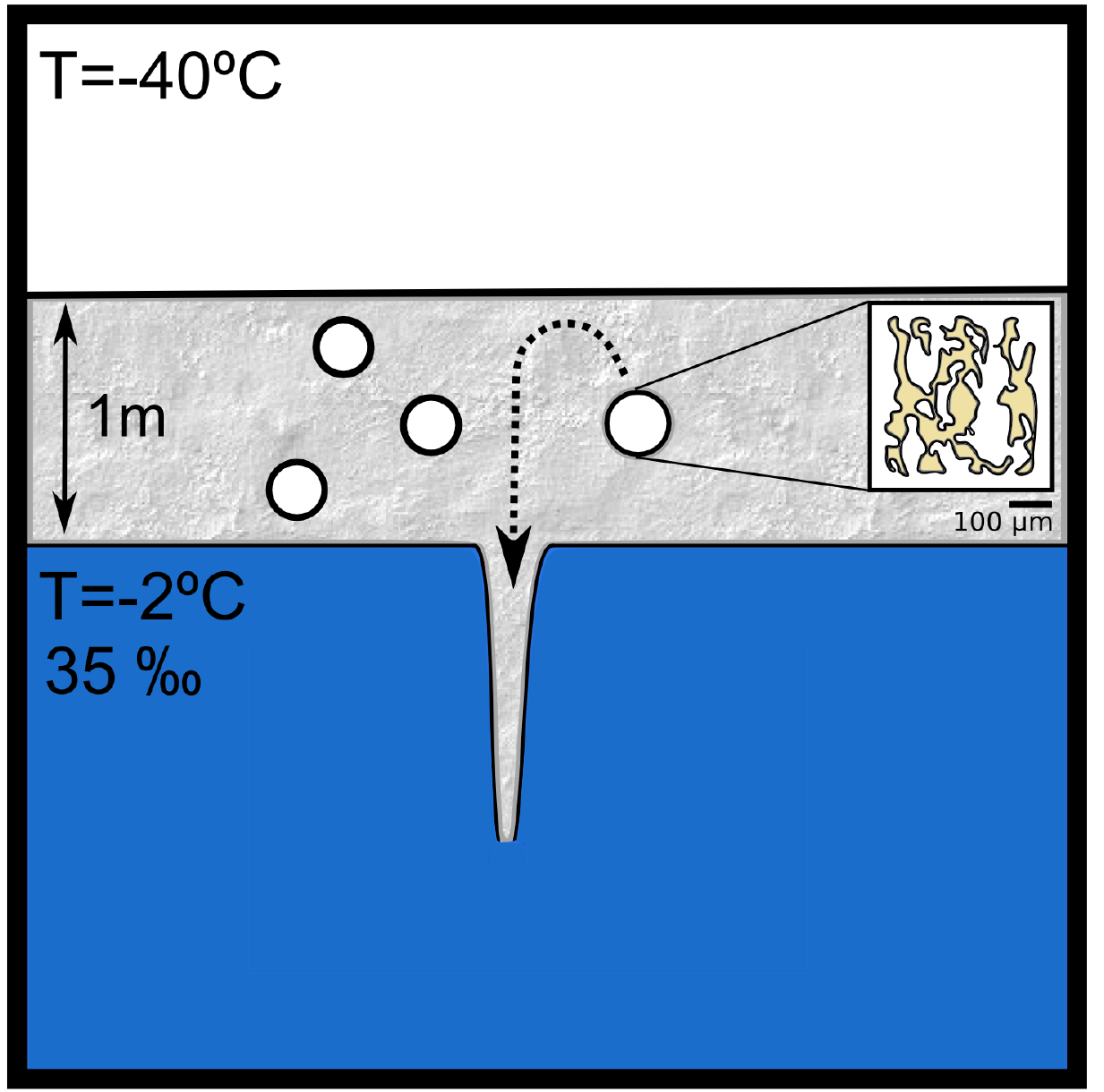}
\end{center}
\caption{Cross section through sea ice illustrating brinicle formation. The temperature in the ice is defined by a gradient between the temperatures of the air and the sea. The brine inside the ice is siphoned through the channel network and ejected through a single opening, forming a tubular brinicle.}
\label{fig:diagram}
\end{figure}

Brinicles are only found during the winter in the polar regions.
As the polar winter progresses, the air temperature above the sea ice drops from -10\textcelsius{} to at least -40\textcelsius{}, while the sea temperature beneath the ice remains at -2\textcelsius{};  see Fig.~\ref{fig:diagram}.
This produces an increasing temperature gradient across the sea ice and heat flows from the sea to the atmosphere.
As the sea water loses heat, ice crystals begin to nucleate in the region directly beneath the ice pack. 
Ice formed in these conditions usually grows in a bi-dimensional way, forming what are referred to as ice platelets \cite{RevModPhys.84.885}.
Because ice is less dense than liquid water, these platelets float upwards and accumulate under the ice pack forming a porous polycrystalline layer known as the skeleton layer \cite{lake1970}.
Within this layer some sea water becomes trapped and, as the heat flow towards the air continues, it begins to freeze, adding its water molecules to the crystal structure that forms the skeleton layer.
As a result the trapped sea water becomes a progressively more concentrated solution of brine.
The trapped brine continues to concentrate and as it does so it becomes denser and colder as it equilibrates with the temperature of the surrounding ice.
At some point, this dense cold brine finds a way to escape from the ice into the sea water below. 
Because brine is denser than the water, it flows downwards and, because it is colder, it absorbs heat from the surrounding water, which is already near its salinity-determined freezing point.
The consequence is that the brine streamer forms a tube of ice, a brinicle, around itself by freezing the surrounding sea water.
The brinicle continues to grow downwards as long as there is brine flowing and as long as there are no strong currents in the sea beneath the ice or movements of the sea ice itself to dislodge it. This growth process was recently filmed \emph{in situ} in Antarctica for the first time in a BBC documentary film, ``Frozen Planet'' \cite{BBC}.

\section{The growth process of brinicles}

The brinicle growth process (Fig.~\ref{fig:diagram}) begins with the formation of the skeleton layer beneath sea ice.
This is a porous polycrystalline mass, like a soaked sponge, that has been termed a mushy layer \cite{Feltham2006}. The characteristic pore size is small enough to inhibit convective fluid exchange with the sea water beneath \cite{yih1959,lake1970}.
This inhibition is a necessary condition for brine entrapment, otherwise the heavier concentrated brine would simply flow into the sea by free convection.
As the ice pack increases in thickness brine gets colder and more concentrated.
When most of the water has frozen, the remaining super-concentrated brine is trapped in a network of so called brine channels.
These compartments are roughly cylindrical, with thin branches with a typical diameter of $\sim$0.1~mm that interconnect with others \cite{Weissenberger,eide1975}.
The development of the brine-channel network is a self-organized process in which
the vertical temperature gradient through the ice favours the formation of compartments where brine remains trapped.

There are several candidate mechanisms for brine migration within the ice pack \cite{untersteiner1967}.
Brine might diffuse from one inclusion to another through the solid ice, but the velocity of surface diffusion is of $\sim1\mu$m$^2$ per hour, which makes it negligible.
Density-driven convection can only occur once the density gradient is strong enough to a provoke instability, which is a function of the radius of the brine channels: $\Delta \rho > f(1/r^4)$ \cite{yih1959}.  
Since the typical radius $r$ of the channels is of $\sim$0.1~mm this mechanism is only effective on the wider channels and only once the ice fracturing has opened enough connections through the channel network. 

The mechanism that starts the growth of a brinicle necessarily implies releasing this trapped brine. A plausible trigger
involves cracks that appear in the ice when the water in the brine inclusions freezes, increasing in volume and building up internal pressure in the crystal lattice.
Cracks in the ice will follow lines of natural weakness, which in the case of a polycrystalline material are the boundaries between crystal grains. 
Those cracks will continue to propagate until they find a hole in the lattice which would stop the fracture.
In the case of sea ice such holes would be the brine channels.
The self-organized structure of the channel network will now guide the fracture along the thinner channels towards wider ones, until it reaches an opening wide enough so that the weight of the brine column is enough to lead to convective instability in the skeleton layer.
It has been found in field observations that one larger channel several millimetres in diameter appeared on average every 180~cm$^2$ \cite{lake1970}.
This sparsity and the idea of fracture propagation help explain why brinicles are often found growing only from the widest brine channels in the surrounding ice \cite{dayton1971}. Such a wide opening is commonly referred to as a `chimney' in mushy layers \cite{Worster1991,Schulze1999,Feltham2006}.

\begin{figure*}[t]
\vspace*{2mm}
\begin{center}
\includegraphics*[width=\textwidth]{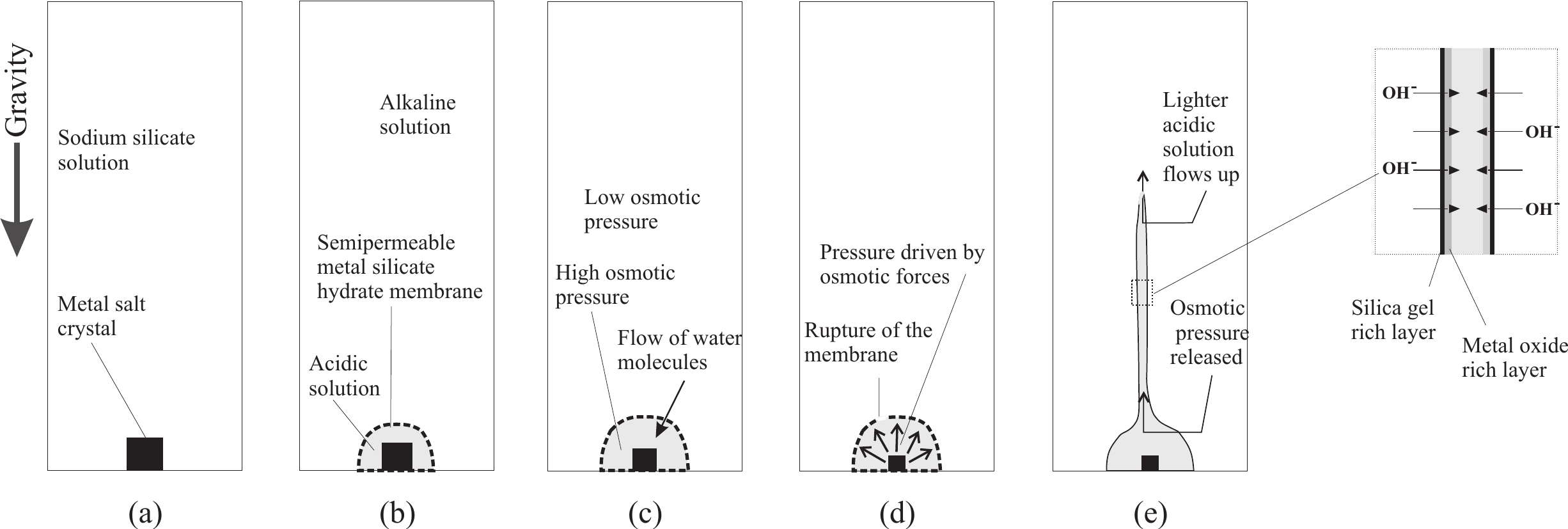}
\end{center}
\caption{\label{fig:mechanism}
Chemical-garden growth:
(a) setup at start of the reaction, (b) membrane formation between acidic and
basic solutions, (c) osmotic pressure is higher within membrane than outside it,
so it expands, (d) under osmotic forces the membrane ruptures, and 
(e) a tube forms.}
\end{figure*}

\begin{figure*}[t]
\vspace*{2mm}
\begin{center}
\includegraphics*[width=\textwidth]{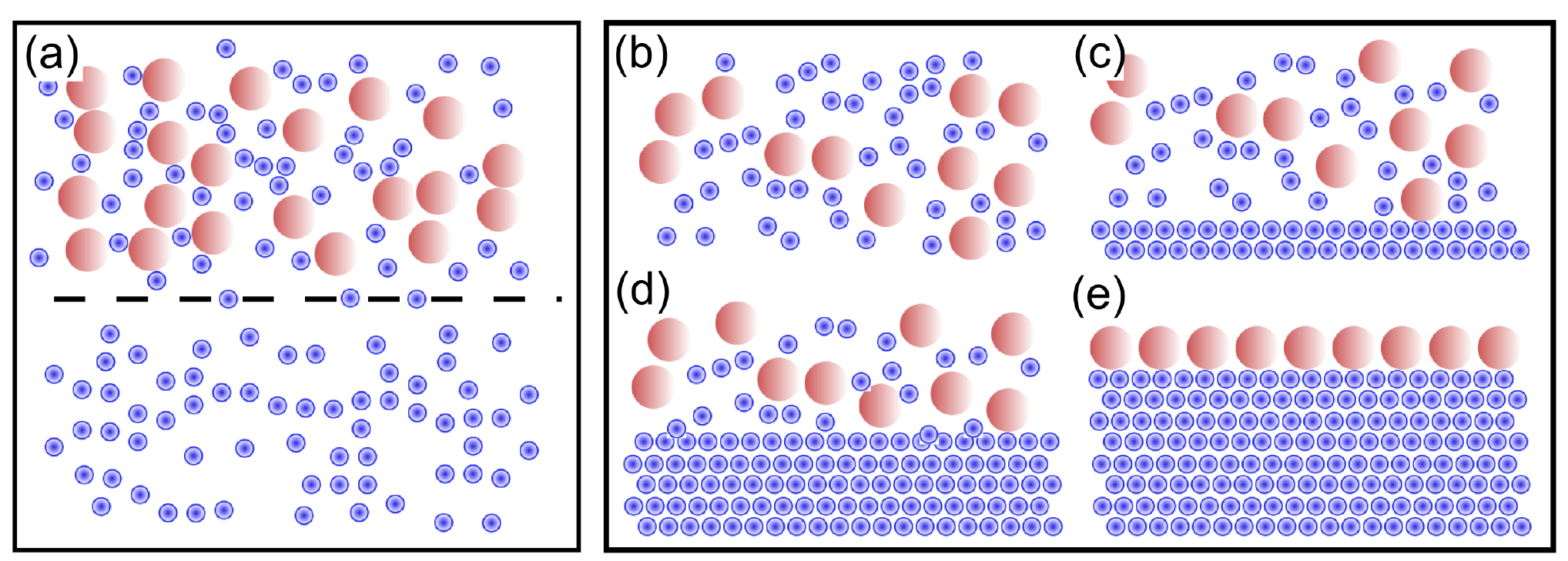}
\end{center}
\caption{Filtering mechanisms: osmosis across a semi-permeable membrane (left) and salt rejection (right).}
\label{fig:relation}
\end{figure*}

Once the concentrated brine has found a way to escape from the ice, it starts flowing into the sea water below and gaining heat by freezing the water around it.
A nearly cylindrical, externally slightly tapering tube then starts to form, growing from the base of the ice downwards into the sea.
This brinicle drains the brine in the surrounding ice pack,  
and produces an inhibiting effect whereby the appearance of one brinicle inhibits the growth of others in the nearby area.
It has been reported that only one major brinicle can be found for every 6--8~m$^2$ of ice \cite{paige1970}.
This inhibition effect is common in other mushy layers, leading to spatial patterns like bi-dimensional square and hexagonal lattices \cite{Worster1997}.

The sea water outside the brinicle is warmer than the brine flowing through the inside so there is heat transfer through the tube walls, freezing more water on the outside and melting it on the inside.
The brinicle thus increases in length and in diameter by ablation of the inner wall and accretion at the outer wall; this explains the slightly conical form of a brinicle.
However, this growth implies that as ice on the inner wall is melting it is reducing its volume and creating a volume deficit inside the tube. 
So there has to be more brine coming into the tube from the ice sheet than there is leaving the tube at the tip.
Furthermore, the fluid closer to the inner wall is warmer and lighter than the brine, meaning that the flow inside the tube can be convectively unstable.
If overturning is produced inside the brinicle then there could be freezing in the brine channels and the growth would be interrupted.
The condition for this not to happen is that the pressure-gradient force inside the tube is always greater than the buoyancy force generated by melting of the inner wall \cite{martin1974}.
For this condition to be satisfied demands that the volume flux be continuous and always above some minimum value, which requires a large amount of brine.

Some brinicles reach lengths of several metres and grow for several hours or days. 
We can estimate how much brine is required to grow such a brinicle.
If we consider the brinicle to be a hollow cylindrical shell, its mass is $\pi(r_2^2 - r_1^2)L\rho_{ice}$, where $L$ is the length of the brinicle, $r_1$ and $r_2$ the inner and outer radii and $\rho_{ice}$ the density of the ice.
Considering  the latent heat of fusion for water, $L_f = 333.7$~J g$^{-1}$, we can estimate how much heat it is necessary to absorb in order to freeze such a mass of ice.
And we can consider how much heat the brine is able to absorb if we suppose that its specific heat is similar to that of water ($c_p \approx 4.18$ J g$^{-1}$K$^{-1}$) 
and that it is flowing into a sea with a temperature difference of $\Delta T\approx20$~K.
With such an estimation we calculate that the total volume of brine necessary to grow a brinicle of $\sim$1~m in length is of the order of $\sim 10^2$ L.
This implies that the flow needs to be $\sim$1 L per minute, which is consistent with field observations \cite{dayton1971}.
If we estimate how much sea water is necessary to generate this volume of brine, considering that sea water has a salinity of $35$\textperthousand\ at $-2$\textcelsius\ and the brine has $224$\textperthousand\ at $-20$\textcelsius, we come up with $\sim10$~L of sea water per litre of brine.
%This implies that a large volume of the ice pack is draining all of its brine through the same brine channel and with maximum efficiency.  One possible way to explain this phenomenon is to think of the brine channel network as a tree with branches, which have developed through the self-organized process of salt rejection.  In this sense it is very similar to a mud volcano.
This amount of liquid will dictate the typical spacing between brinicles.

\section{Brinicles as chemical gardens}

The formation process of a chemical garden is described in detail in Figure~\ref{fig:mechanism}.
The setup for the reaction is a metal salt crystal immersed in a sodium silicate solution (Fig.~\ref{fig:mechanism}a).
As soon as the seed crystal comes into contact with the aqueous solution, it begins to dissolve and at the same time is covered by a colloidal
coating (Fig.~\ref{fig:mechanism}b). 
This material acts as a semipermeable membrane. 
As a consequence of the different osmotic pressures inside and outside the membrane, water is drawn osmotically from the outside,
permitting the further dissolution of the crystal (Fig.~\ref{fig:mechanism}c). 
The entry of water causes the membrane to dilate under osmotic pressure until it ruptures (Fig.~\ref{fig:mechanism}d). 
This provokes the injection of the salt solution in its interior into the silicate solution. 
As the jet of solution is generally  lighter than the fluid around it, it flows upwards by buoyancy and an upwards-pointing tube grows around the expelled jet  (Fig.~\ref{fig:mechanism}e). However, chemical gardens can also grow downwards under conditions in which the density difference is reversed and  the jet is denser than the surrounding liquid \cite{Jones1998,Pagano2007}.  Such downwards growing tubes have been termed reverse or inverse chemical gardens.

At this point, to compare 
the filtering mechanisms of chemical gardens and brinicles, it is important to consider the process of salt rejection by sea ice and to compare it to the semipermeable membrane of a chemical garden filters out solute particles that are too large to go through its pores and allows only water molecules to pass; Fig.~\ref{fig:relation}.
The polarity of water molecules determines that they will crystallize into a hexagonal crystal lattice \cite{petrenko}.
This is crucial for our planet because hexagonal ice is less dense than liquid water and hence floats on the sea. 
Without this unusual property of water our ocean would freeze and life as we know it would not be possible.
But the hexagonal crystal lattice also plays an important role in salt rejection of sea ice.
When water molecules start to crystallize, everything that does not enter into the lattice is pushed away from the solid--liquid interface.
Hence this is a process of purification at the molecular level, very much like the filtration process through a semipermeable membrane but in an inverse sense.
In a classical chemical garden there are two solutions with different concentrations separated by a semipermeable barrier with a flow of water from smaller to greater concentration, which provides energy for an osmotic pump. A similar osmotic pump mechanism makes brinicles equivalent to chemical gardens: The higher osmotic pressure of the concentrated brine in the mushy layer will cause an inflow of water, and it is this continual inflow --- continual for as long as the osmotic pressure imbalance remains owing to salt rejection during crystallization --- that maintains the osmotic pump of the brinicle.

The growth of chemical gardens grown on Earth is driven by a combination of the potential energies from an osmotic pump and density-driven convection  \cite{cartwright}. Likewise, density-driven convection as well as osmosis is important in brinicles; only that with a brinicle, the density difference is reversed from a classical chemical garden, so the brinicle grows downwards. We may view this convection through the brinicle as a form of siphon: as long as the brinicle's tip is at a lower height than the bottom of the ice sheet the brine will continue to flow and, considering Bernoulli's equation, the flow will be nearly constant (see Figure \ref{fig:diagram}). Seeing the brinicle system as a siphon solves the problem of explaining how the flow remains constant for the long times required for brinicle growth. It also explains where all the brine comes from, because one open channel can siphon the brine out of all surrounding channels that are connected through the network, even if they are at a lower 
height than the siphon's neck.
Furthermore, if one open channel can siphon all the brine from several square metres around then it justifies the inhibition effect producing the sparsity reported in field observations \cite{paige1970,dayton1971}.
Density-driven convection in a chemical garden may also be considered a siphon in which the fluid being siphoned is less dense than the surrounding fluid and so the siphon there operates upwards, in reverse.
This siphon effect disappears when growing chemical gardens in microgravity \cite{chemgarden2,chemgarden3}, when the osmotic force alone drives growth, and it would likewise disappear if brinicles were grown in the absence of gravity.

\section{Conclusions}

The formation process of brinicles remains little studied because of the difficulties of field observations:
They grow only in polar regions, in calm waters and under the ice sheet. In common with other unusual types of ice formed under polar conditions, such as anchor ice \cite{Denny01102011}, there is not a great deal of published work and there are remain many questions to answer. Here we have placed brinicles within the framework of chemical gardens. 
Future work should include quantitative modelling and theory as well as laboratory experiments, in both brinicles in particular and chemical gardens in general, as part of this new research area of {\it chemobrionics} \cite{barge2013}.

%We have used the knowledge accumulated in the field of chemical gardens to understand better brinicle formation and we have proposed a plausible explanation for most unexplained questions. Future work should include laboratory experiments and dynamical simulations.

Like some other tubular formations in geology, brinicles can be understood as a form of chemical garden; in this case an inverse chemical garden.
The crystal lattice of sea ice can filter out water molecules and accumulate highly concentrated brine in the opposite way in which an osmotic membrane can filter out the solute and accumulate water molecules in the case of chemical gardens.
Once the filtering and concentrating is done, the resulting brine solution will be expelled into the sea and the growth of a brinicle begins by freezing water around the brine streamer. 
In the case of chemical gardens a similar hollow tube is formed --- generally growing upwards in this case --- by the precipitation of metal silicate around a jet of metal-salt solution.
But both freezing and precipitation can be regarded as parallel processes if we consider that both systems are losing energy to become more stable. Both osmosis and density-driven convection contribute to the self-organized formation of self-assembled tubular precipitation structures.

Understanding the formation of brinicles goes hand in hand with understanding the process of salt rejection in sea ice.
The concept that salt rejection works in a similar way to semipermeable membranes is an idea that can have several implications in many related processes.
Reverse osmosis is currently used in desalination plants \cite{Fritzmann2007} to achieve basically the same effect that is achieved in sea ice: purifying salt water. 
The rejection of impurities in sea ice is also very similar to industrial processes related with metallurgy, where impurities in the solidifying metal can affect its mechanical properties \cite{WETTLAUFER}.
It also has medical implications in the fields of bio-compatible materials \cite{Deville27012006,Wegst28042010} and in the controlled freezing of biological tissues \cite{rall1985}.

But perhaps the most important application of salt rejection is that related with the theories for a cold origin of life on our planet or elsewhere in the universe.
The origin of life is often proposed to have occurred in a hot environment, like the one found in hydrothermal vents \cite{corliss1979}. It is proposed that chemical-garden processes are involved in the mechanism \cite{russell,russell2007}.
But there is a different school of thought that presents sea ice as a promoter of the emergence of the first life \cite{trinks2005,RevModPhys.84.885}.
Brine rejection in sea ice produces all the conditions that are considered necessary for life to appear.
We have mentioned the enrichment of chemical compounds that occurs in brine entrapment between ice-crystal grains, but there is also membrane formation by deposition of lipids, today originating from extracellular polymeric substances (EPS) generated by phytoplankton \cite{Krembs01032011}, and perhaps produced by complex prebiotic molecules at the dawn of life.
There are electric potentials and pH gradients across the interface of ice and brine, and the surface of ice has been proven to have catalytic effects \cite{trinks2005}.
As brinicles play an important role in the dynamics of brine transport through sea ice, they might also play a role in this scenario  of  a cold origin of life, just as hydrothermal vents do in the hot environment theories, and in both instances chemical-garden processes are fundamental.

Beyond Earth, the brinicle formation mechanism may be important in the context of planets and moons with ice-covered oceans. As well as similar brinicles to those on Earth, under astrophysical conditions of different temperatures and pressures, there exists also the possibility of brinicles forming from other ice phases beyond the familiar hexagonal ice Ih, including high-pressure ice phases that are more dense than water \cite{RevModPhys.84.885}, so that such a brinicle, say on Ganymede or Callisto, might grow upwards from the ocean floor, rather than downwards from above. One might speculate that brinicles might play a similar role on such icy bodies as hydrothermal vents are proposed to have played in the origin of life on Earth.

%%%%%%%%%%%%%%%%%%%%%%%%%%%%%%%%%%%%%%%%%%%%%%%%%%%%%%%%%%%%%%%%%%%%%
%% The "Acknowledgement" section can be given in all manuscript
%% classes.  Rather than use \section, an appropriate macro is
%% provided that will always work.
%%%%%%%%%%%%%%%%%%%%%%%%%%%%%%%%%%%%%%%%%%%%%%%%%%%%%%%%%%%%%%%%%%%%%
\acknowledgement
IT acknowledges the financial support of the Spanish Ministerio de Ciencia y Innovaci\'on grant FIS2010-22322-C01 and a Ram\'on y Cajal fellowship.
JHEC and CISD acknowledge the financial support of the Spanish Ministerio de Ciencia y Innovaci\'on grant FIS2010-22322-C02. This work is published within the framework of the NASA Astrobiological Institute focus group on thermodynamics, disequilibrium and evolution (TDE).

%%%%%%%%%%%%%%%%%%%%%%%%%%%%%%%%%%%%%%%%%%%%%%%%%%%%%%%%%%%%%%%%%%%%%
%% The appropriate \bibliography command should be placed here.
%% Notice that the class file automatically sets \bibliographystyle
%% and also names the section correctly.
%%%%%%%%%%%%%%%%%%%%%%%%%%%%%%%%%%%%%%%%%%%%%%%%%%%%%%%%%%%%%%%%%%%%%

%\bibliographystyle{biochem}
\bibstyle{biochem}
\bibliography{brinicles}

%\tocentry
%\begin{tocentry}
%   \includegraphics[height=4cm]{TOC_graphic.pdf}
%   Schematic illustration of brinicle formation.
%\end{tocentry}

\end{document}